\documentclass{article}

\usepackage{PRIMEarxiv}

\usepackage[utf8]{inputenc} 
\usepackage[T1]{fontenc}    
\usepackage{hyperref}       
\usepackage{url}            
\usepackage{booktabs}       
\usepackage{amssymb}        
\usepackage{nicefrac}       
\usepackage{microtype}      
\usepackage{lipsum}
\usepackage{fancyhdr}       
\usepackage{graphicx}       
\graphicspath{{media/}}     

\title{A possible solution to the mystery of the ANITA anomalous events
}

\author{
  Massimo Villata \\
  INAF, Osservatorio Astrofisico di Torino, I-10025 Pino Torinese (TO), Italy \\
  \texttt{massimo.villata@inaf.it} \\
}

\begin{document}
\maketitle

\begin{abstract}

In 2006 and 2014, the Antarctic Impulsive Transient Antenna (ANITA), a balloon-borne radio observatory flying over Antarctica, detected two strange upward-going radio pulse events that have not yet been explained by our current understanding of physics. These were not signals reflected by the ice and therefore it must have been an air shower originating from a cosmic ray coming from under the Antarctic ice, but this hypothesis was also ruled out by various data analyses. The CPT gravity theory and its associated cosmological model, the lattice Universe, can instead explain those events in a completely natural and spontaneous way, without any additional assumptions beyond general relativity and the expected matter-antimatter symmetry of the Universe on which they are based. Together with the antihelium candidate events from AMS-02, the anomalous ANITA events can thus lend further validity to a cosmological model that has already achieved considerable success in explaining the accelerated expansion of the Universe, without the need for dark energy. These events thus add to a series of problems unsolved by standard cosmology and physics, but whose solution is straightforward, spontaneous and natural within the framework of CPT gravity, without the need for ad hoc hypotheses and unknown ingredients.

\end{abstract}


\section{Introduction}

The Antarctic Impulsive Transient Antenna (ANITA) experiment, a balloon-borne radio observatory launched by NASA, was designed to detect ultrahigh-energy neutrinos through the Askaryan effect in ice.\cite{AniCol2009,Hoo2010} It flew at $30$--$39\rm\,km$ altitudes above Antarctica, and detected radio pulses that are consistent with coherent emission from ultrahigh-energy cosmic-ray (UHECR) air showers. These pulses showed large horizontal polarization, as a consequence of the nearly vertical Earth’s magnetic field.

Conventional descending UHECR air showers produce downward-propagating radio impulses that ANITA could detect in reflection off the ice surface, thus appearing as upward-going events arriving from below the horizon. In addition, they exhibited the characteristic polarity inversion due to reflection. These events represented the great majority of the ANITA cosmic-ray-like detections, and could be easily explained.\cite{Gor2016,Gor2018,Gor2021}

ANITA detections of UHECR events also include a subset of near-horizontally-propagating stratospheric air showers arriving from just above the horizon, which point directly at the instruments, without reflection off the ice and thus with no polarity inversion of the signal. Even a few trajectories coming from slightly below the horizon can be interpreted as unreflected signals, if there is no phase inversion, and if the path through the Earth to be traversed is as short as to allow the transit of the supposed ultrahigh-energy neutrino without absorption.

Surprisingly, ANITA reported two upward-going events, one during the first flight in 2006 (ANITA-I)\cite{Gor2016} and the other during the third flight in 2014 (ANITA-III),\cite{Gor2018} at payload arrival angles of $-27^\circ$ and $-35^\circ$ relative to horizontal, i.e., well below the horizon, but without the polarity inversion expected from ice reflection. These detections defie our current physical knowledge, because the direction of the observed pulses implies that the expected tau neutrinos originating them would need to travel around $6000$--$7000\rm\,km$ through the Earth, which corresponds to about $8$--$10$ interaction lengths at the required neutrino energy $E_\nu\gtrsim0.2\rm\,EeV$\cite{Con2011} ($1{\rm\,EeV}=10^{18}\rm\,eV$). The consequent strong attenuation would require a tau-neutrino flux that should have been observed with IceCube and the Pierre Auger Observatory (e.g., ref.\ \cite{Abd2025} and references therein).

Several attempts have been made to explain these anomalous events observed by ANITA, both with standard physics and with exotic concepts and particles. None of these explanations has seemed convincing, especially after comparison with the observations and data analyses of the two other observatories (IceCube and Pierre Auger) capable of competing with ANITA in this specific context. Their results seem to indicate a disagreement with the interpretation of the unusual events as upward-going showers, produced by neutrinos or any other particle traveling through the Earth (e.g., ref.\ \cite{Abd2025} and references therein).

Now, the question we would like to answer is: is there an already existing theory, even if unconventional, that can explain these strange events detected by ANITA? Yes, there is, and the mysterious events could have been predicted even before they were observed.

\section{CPT gravity and time reversal}
\label{sec:cpt}

Gravity appears to be always attractive, at least on Earth, in the Solar System, in the Galaxy and its immediate extragalactic neighborhood. At larger scales, well-known problems arise. The Local Sheet already exhibits anomalies that cannot be explained by purely attractive gravity, and the problem worsens further when we consider the general expansion of the Universe and the recent discovery of its acceleration over time.

Our current best understanding of gravitation comes from general relativity, which seems almost infallible, at least in the aforementioned neighborhood. So much so that efforts have been made to make it compatible even with the accelerated expansion, by appropriately equipping its cosmological equations with the ad hoc ingredients needed to make the cosmological model match the observations.

It does not seem like a compelling solution; it is a bit reminiscent of Ptolemaic epicycles used to explain planetary motions.

But let us start with general relativity anyway and see if there is room for maneuver to interpret the incriminated phenomena more elegantly.

General relativity was formulated well before the discovery of the existence of antimatter. The latter also presents another well-known problem: in our Universe, antimatter appears to be almost absent, while one would expect a symmetry between matter and antimatter. One suspects that the two problems may be connected and that they can be solved together.

What does general relativity say if we replace matter with antimatter in its gravitational equations, using the charge conjugation operation, C? Nothing, because this operation does not affect any of their elements. Someone might think that it could instead have an effect on gravitational mass, changing its sign, and therefore implying a gravitational repulsion between matter and antimatter (just think of Newton's equation). But that would be another ad hoc hypothesis, and instead we want to be purists and fully respect Einstein's work, where mass is a scalar, that is, a relativistic invariant that cannot change its sign or value. Moreover, this has recently been (almost) demonstrated by the ALPHA-g experiment at CERN.\cite{alpha2023}

There is another transformation that is often associated with charge conjugation, due to the so-called CPT symmetry, that is, the invariance of physical laws when C, P (the parity operation, which changes the sign of the spatial coordinates), and T (the time reversal, which inverts time and therefore the sequence of events), are simultaneously applied to a physical system. Note that the PT transformation is a proper Lorentz transformation (total inversion) and therefore legitimately applicable to any relativistic system.

Let us check what happens, for example, to the geodesic equation:
\begin{equation}
\label{eq.1}
{{\rm d}^2x^\alpha\over{\rm d}\tau^2}=-{\Gamma^\alpha}_{\beta\gamma}{{\rm d}x^\beta\over{\rm d}\tau}{{\rm d}x^\gamma\over{\rm d}\tau}\,,
\end{equation}
which is composed of four elements, three four-vectors, and the affine connection ${\Gamma^\alpha}_{\beta\gamma}$. Nothing happens, as expected from CPT invariance. In particular, as we already know, C has no effect, while PT changes the sign of each of the three four-vectors and that of the affine connection: four sign changes, so the equation remains unchanged. But if C does not affect the equation and this is shown to be invariant under PT, why involve C as well? Because this also guarantees invariance for all other non-gravitational interactions. In any case, what follows from this invariance is that in a system in which matter is replaced by antimatter and space-time is reversed, everything works as before, that is, gravity is still attractive, unchanged.

So, if such a system exists somewhere in the Universe, it behaves like the original one. Inside it, everything happens in the usual way: matter is matter, spatial coordinates are not reversed, and time flows forward, not backward. It is only from the outside that the CPT transformation would be perceived. And our system would be perceived from there as CPT-transformed. It is just a matter of ``perspective'', not of real change. Like when we say that the usual Lorentz transformations dilate time (and contract lengths): for each of the two reference frames, it is the other that exhibits time dilation (and length contraction).

But let us return to our geodesic equation. What would happen to the gravitational interaction between the above two systems separated by a CPT transformation? If we consider a normal test particle interacting with a CPT-transformed affine connection, this results in a sign change in the equation, which signifies gravitational repulsion (i.e., the test particle ``feels'' an inverted gravitational field). And vice versa, if it is the test particle that is CPT-transformed, there will be three changes of sign in the equation, one for each four-vector, and therefore we will have the same repulsive gravity. In conclusion, two opposite space-times gravitationally repel each other.\cite{vil2011,vil2015,Vil2024}

We are almost there. If the Universe were composed of opposite space-times that repel each other and populated by equal amounts of matter and ``antimatter'', we could have solved two problems at once: accelerated expansion and matter-antimatter asymmetry.

But where could these ``islands'' of antimatter, which are supposed to alternate with the ``islands'' of matter we know, be located in the Universe around us? Excellent candidates are the so-called cosmic voids, regions with very little luminous matter, whose underdensity is often invoked as the cause of a kind of gravitational repulsion due to the density contrast between them and the dense regions of luminous matter. But this apparent repulsion is not enough to explain neither the dynamical and morphological anomalies of our Local Sheet bordering the enormous Local Void,\cite{Pee2010} nor, even less, the acceleration of cosmic expansion.

If we supply such ``voids'' with an adequate population of antimatter to balance the amount of known matter, we obtain a good explanation of the above local peculiarities,\cite{vil2012b} and an excellent agreement between the predictions of such a cosmological model, called the ``lattice Universe'', with observations, all without any need for dark energy.\cite{vil2013}

It therefore seems like a completely viable and worthy path, and we will not mention here all the advantages it represents with respect to the unsolved problems of standard cosmology, since these have been pointed out elsewhere.\cite{vil2012b,vil2013,Vil2024}

Perhaps there remains a single, seemingly major problem, or at least that is how some consider it: if voids are populated by antimatter, which is actually just matter in a space-time reversed compared to our own, why do not we perceive it in any way, except through its gravitational repulsion? Why do we see darkness where there should be luminous objects like ours?

The simplest answer is also the one that holds the key to unraveling the mystery of the ANITA anomalous events.

Consider a cosmic void containing its own quantity of matter, CPT-transformed relative to us. This matter can emit particles. Once these emissions pass from their domain to ours, they may encounter Earth, which from their ``point of view'' is an anti-Earth in a space-time inverted with respect to theirs. Here they will meet something capable of absorbing them, and that will be the end of their journey. If it is a photon, it could also reach a detector, such as a CCD camera or a photographic plate, or even a human eye. But how will this event be experienced from our point of view? Because of the particle CPT transformation relative to our space-time, it, coming from its past, will appear to come from our future and apparently endowed with negative energy (which in its space-time is instead rigorously positive). In practice, if we do not have negative-energy detectors, not only would we not ``see'' the photon, but for us it would still travel the reverse path, from its absorber to its source. This can be the explanation for the invisibility of antimatter belonging to an opposite space-time.\cite{Vil2024}

Finally, consider the journey of a high-energy cosmic ray coming from a cosmic void with its opposite space-time. Arriving in the vicinity of the Earth (which for it is CPT-transformed), it hits a component of its anti-atmosphere and gives rise to an air shower, which emits its downward-going radio pulses with non-inverted polarity. These pulses would be detected by us in a reversed time sequence, thus appearing as upward-going events, just like the anomalous ANITA events. But this sequence does not at all require that there be a high-energy particle, such as the hypothesized tau neutrino, which would emerge from under the Antarctic ice after having traveled too long a path inside the Earth. And the mystery seems solved.

Once these upward-going radio pulses are detected, we can imagine that the time reversal would immediately cause the shower to recompact itself and give rise to the antiparticle of the one that produced it, which would then head undisturbed towards its source. And all this should not be surprising, as it is consistent with the Feynman-St\"uckelberg interpretation that antiparticles are nothing else than the corresponding particles traveling backwards in time.\cite{Stu1942,Fey1948,Fey1949} Like a positron in a Feynman diagram, which can be interpreted as an electron taking a reverse time path, coming from our future or returning to its past.

\section{Possible association between anomalous events and nearby cosmic voids} 

Once accelerated by their source, UHECRs can interact with ambient baryonic matter and radiation fields, producing secondary neutrinos. These particles, being electrically neutral, travel in geodesics unaffected by extragalactic and Galactic magnetic fields, and therefore their direction of arrival can precisely indicate the birthplace of their progenitors and the position of the accelerator in the sky (e.g., refs.\ \cite{Kot2011,AlvBat2019}).

As discussed in the previous section, in our scenario this position is not to be found in the apparent arrival direction of the radio pulses, i.e., the one below the horizon that represents an impossible path even for high-energy neutrinos, but in the opposite direction, the upward-going one, in which the radio pulses are directed from our point of view. If the cosmic ray coming from there is a secondary neutrino produced near the source, then the location of the source can be pinpointed very precisely. Conversely, if the cosmic ray arriving on Earth is a charged particle (such as a proton or a heavier nucleus), then it will be necessary to consider that the aforementioned position, due to the deflections undergone by the particle in magnetic fields, will have some uncertainty, which increases with the charge of the particle and decreases with its energy, and may even reach several degrees (e.g., refs.\ \cite{Kot2011,AlvBat2019} and references therein).

\begin{figure}
\centering
\includegraphics[width=15.5cm]{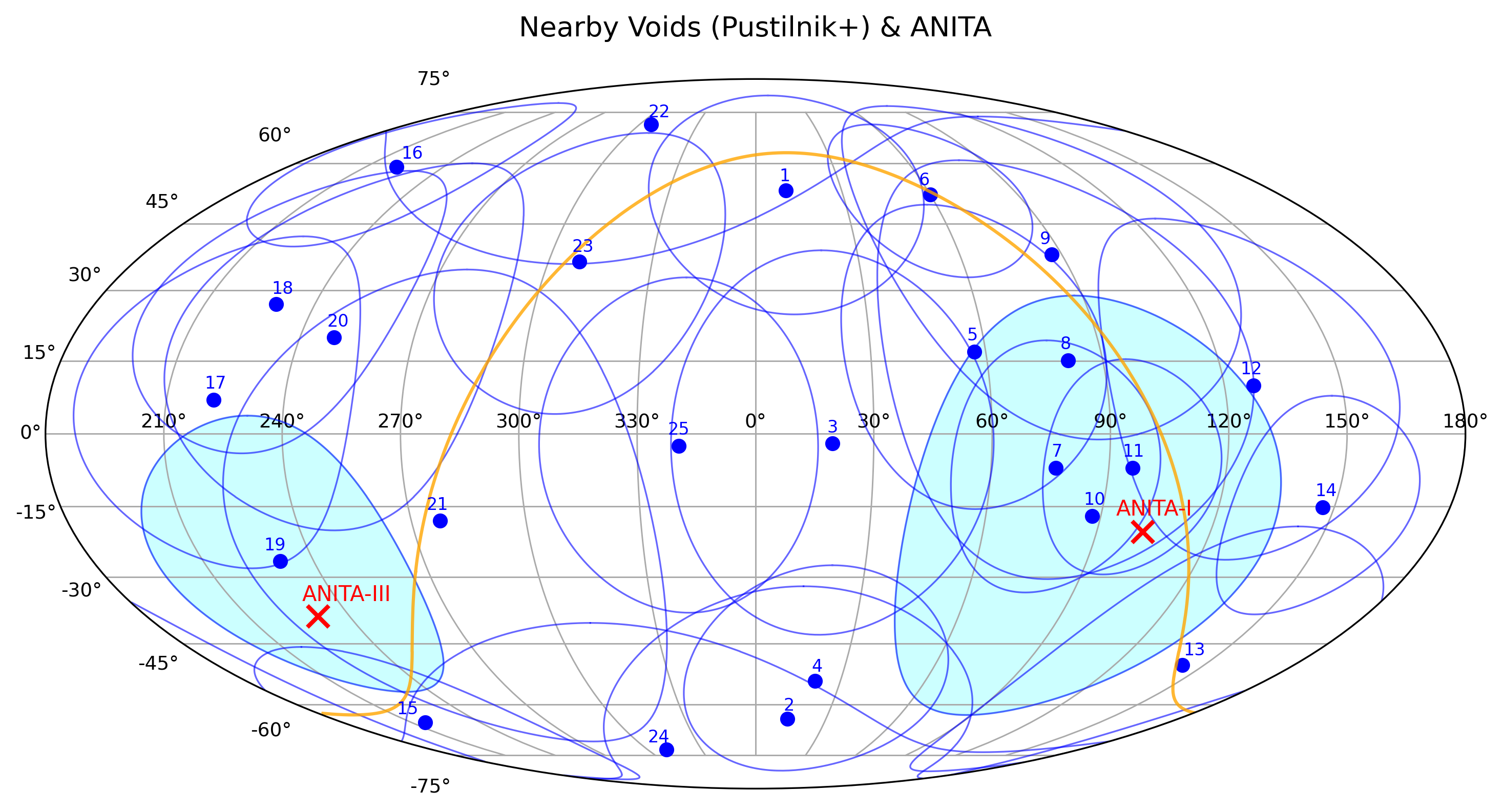}
\caption{Sky distribution in equatorial coordinates of the nearby cosmic voids listed in Table 1 of ref.\ \cite{Pus2019}. The blue dots mark the centers of the voids, and the blue circles represent their approximate extent in the sky. The red crosses are the positions of the upward-going directions of the two ANITA anomalous events. The cyan shaded areas are the likely host voids of their sources. The orange curve indicates the Galactic plane.\label{fig1}}
\end{figure}

In Figure \ref{fig1} we show the sky distribution in equatorial coordinates of the nearby cosmic voids listed in Table 1 of ref.\ \cite{Pus2019}. Superimposed as red crosses are the arrival directions of the two anomalous events, ANITA-I and ANITA-III, that is, the directions in which the detected upward-going radio pulses point.

The ANITA-I event is located well inside Void \#10 (Lepus), which looms large on the sky, especially because it is very close to us, with its center estimated at a distance of $6.3\rm\,Mpc$ and a roughly equal radius. Other voids around there are at various distances and just graze or touch the position of ANITA-I, reducing the likelihood that they could be the hosts of the source.

ANITA-III is positioned well within Void \#19 (Libra, which, with its center at about $18\rm\,Mpc$, is not particularly close to us), and barely intercepts the enormous and very nearby Void \#21 (Ophiuhus-Sagittarius-Capricornus), which includes the near part of the gigantic Local Void, bordering our Local Sheet. Mainly because of the uncertainty in the position in the case of a charged cosmic ray, it is difficult to say which of the two might be the most likely host for the ANITA-III source.

This correspondence of the ANITA directions with the nearest cosmic voids is not a proof of the validity of the lattice Universe model and the theory of CPT gravity. This validity has been tested in comparison with observational cosmological data and in its ability to solve problems that standard cosmology cannot explain without adopting ad hoc hypotheses. These results will be briefly and partially recalled in the last section.

The aforementioned correspondence instead represents a useful clue in favor of the identification of cosmic voids as locations of space-time opposite to our own. This evidence should be verified with future detections of events similar to those of ANITA, by new and more sensitive instruments in preparation, such as those of the upcoming Payload for Ultrahigh Energy Observations (PUEO) mission.

A further and formidable clue in this sense could come from the possible confirmation of the detections of antihelium nuclei by the Alpha Magnetic Spectrometer (AMS-02).

\section{Antihelium nuclei candidate events} 

The AMS-02 collaboration reported observation of eight antihelium candidate events. Six of them have mass in the range of antihelium-3, and two candidates have mass in the range of antihelium-4. Although these results have not yet been published in a peer-reviewed journal, should they be confirmed, the detection of cosmic antihelium would be a breakthrough discovery with profound implications for our current understanding of cosmology. The discovery of even a single antihelium-4 nucleus is challenging to explain in terms of standard physics, and was long known as being a ``smoking gun'' signature of the presence of anti-objects in the cosmos, in particular anti-stars, and some investigations have been conducted in this context (e.g., refs.\ \cite{vonDoe2020,Byk2023} and references therein).

In contrast to the difficulties of explaining the presence of antihelium both in known physics and in the hypothesis of Galactic or extragalactic anti-objects, in our scenario, the existence of antihelium is natural and physiological, expected and inevitable, given the presence of space-times identical but CPT-opposite to ours, which would certainly have had their own primordial nucleosynthesis and nuclear fusion inside anti-stars, as well as the necessary cosmic-ray accelerators. As with the other cosmic rays we discussed in connection with the ANITA anomalous events, the antihelium nuclei would come from the nearby voids shown in Figure \ref{fig1}, most likely from those bordering the Local Sheet.

\section{Discussion and conclusions}

Why should there be opposite space-times in the Universe? If we believe in the hypothesis that the Universe was born with some ``big bang'' from a no-space-time state, then it seems more likely that, by some sort of conservation law, this gave rise to opposite space-times; a bit like the creation of particle-antiparticle pairs from a photon energy that in itself has nothing to do with space and time, since for the photon ${\rm d}s={\rm d}\tau=0$. The so-called big bang would not be an ``explosion'', nor would it need a bizarre and almost instantaneous inflationary period to solve the horizon problem, because the expansion of the cosmos, and its acceleration, can be explained by the gravitational repulsion between opposite domains, which also implies that a horizon problem does not exist.\cite{vil2013,Vil2024}

This gravitational repulsion, consistent with general relativity, also avoids the problem of dark energy, which looks very much like a mere ad hoc mathematical hypothesis, without any physical meaning. The lattice Universe model built on the theory of CPT gravity can explain a wealth of observational evidence, at least as well as the dark energy model, and can solve problems unsolved by standard cosmology. In particular, it can account for the dynamical and structural anomalies of the Local Sheet bordering the Local Void, once this and the other cosmic voids are endowed with the necessary amount of ``antimatter'' to re-establish symmetry with matter.

This model also implies that the Universe may be older than the about $13.8\rm\,Gyr$ of the standard model, up to approximately $15\rm\,Gyr$, thus allowing more time for the formation of the oldest stars that would otherwise have occurred too close to the big bang. This could also explain the recent discoveries of numerous bright, massive, and ``mature'' galaxies that existed just a few hundred million years after the big bang, thus challenging existing models of galaxy formation, which predict that galaxies of this size would take billions of years to form. Alternatively, or complementary, we know that the repulsive gravity of cosmic voids can help a lot to rapidly gather matter into galaxies and structures, especially at the edges of voids, thus accelerating galaxy formation.\cite{vil2012b,vil2013}

On a cosmological scale, the lattice Universe successfully passes the test of the Hubble diagram of Type Ia supernovae, just as well as the $\Lambda$CDM model, and in its cosmological equations it displays a repulsive energy entirely comparable to that of dark energy, with the difference that the repulsion is not due to a uniform negative pressure in space, but to discrete repulsors in the cosmic voids.\cite{vil2013} And this discrete distribution can also explain dark flows and other excessive inhomogeneities and anisotropies of the Universe, as well as unexplained kinematics on medium and small scales. The repulsive effect of the cosmic voids surrounding galaxy clusters can contribute to (or be the cause of) their confinement, making the potential well shallower at the center, which can also explain the existence of the wobbling of the brightest cluster galaxies, which instead should be kept tightly bound at the center of the cluster by the cuspy density profile of cold dark matter.\cite{Har2017}

Finally, as we have shown in this paper, the same lattice Universe model can provide a natural and predictable solution to the mystery of the anomalous ANITA events, and a spontaneous explanation of the antihelium candidate detections by AMS-02.

All this with a single, coherent cosmological model, consistent with CPT symmetry and general relativity, in agreement with the expected matter-antimatter symmetry, with no ad hoc hypotheses or unknown ingredients.


\bibliographystyle{unsrt}  
\bibliography{references}

\end{document}